\documentclass[10pt,oneside,twocolumn,aps,pra,preprintnumbers,bibnotes,superscriptaddress,amsmath,amssymb]{revtex4-2}

\usepackage[utf8]{inputenc}
\usepackage{graphicx} 
\usepackage{subfig}
\usepackage{caption}
\usepackage{subcaption}
\usepackage{epstopdf}

\usepackage{amsmath}
\usepackage{amssymb}
\usepackage{amsfonts}
\usepackage{amsthm}
\usepackage{mathtools}
\usepackage{bm}
\usepackage{braket}
\usepackage{physics}
\usepackage{cancel}
\usepackage{float}

\usepackage[usenames,dvipsnames]{xcolor}
\usepackage{ulem}

\usepackage{hyperref}
\usepackage{xurl}
\usepackage{lipsum}
\usepackage{comment}

\usepackage{tikz}
\usepackage{tikzscale}
\usetikzlibrary{quantikz2,calc,arrows,chains,matrix,positioning,scopes}

\usepackage{letltxmacro}
\LetLtxMacro{\originaleqref}{\eqref}
\renewcommand{\eqref}{Eq.~\originaleqref}

\begin{document}

\title{Quantum work statistics and coherence effects in quenched bosonic Josephson junctions}

\author{Mattia Orlandini}
\affiliation{Istituto Nazionale di Ottica del Consiglio Nazionale delle Ricerche (CNR-INO), Largo Enrico Fermi 6, I-50125 Firenze, Italy}

\author{Stefano Gherardini}
\affiliation{Istituto Nazionale di Ottica del Consiglio Nazionale delle Ricerche (CNR-INO), Largo Enrico Fermi 6, I-50125 Firenze, Italy}
\affiliation{European Laboratory for Non-linear Spectroscopy, Università di Firenze, I-50019 Sesto Fiorentino, Italy}

\author{Lorenzo Buffoni}
\affiliation{Department of Physics and Astronomy, University of Florence, 50019 Sesto Fiorentino, Italy}

\author{Beatrice Donelli}
\affiliation{Istituto Nazionale di Ottica del Consiglio Nazionale delle Ricerche (CNR-INO), Largo Enrico Fermi 6, I-50125 Firenze, Italy}
\affiliation{European Laboratory for Non-linear Spectroscopy, Università di Firenze, I-50019 Sesto Fiorentino, Italy}

\begin{abstract}
We investigate the non-equilibrium work statistics originating from a sudden quench in a bosonic Josephson junction. In particular, by employing the Holstein-Primakoff approximation, the work statistics are analytically characterized in the weak-interaction regime, where the dynamics map onto a time-dependent quantum harmonic oscillator. For a junction initialized in the ground state of the pre-quench Hamiltonian, we demonstrate that the work statistics are governed by a negative binomial distribution, as occurs in fully-connected models driven across a critical point. Furthermore, we also consider initial superposition states containing quantum coherences in the energy basis. To characterize the corresponding work distributions, we employ Kirkwood-Dirac quasiprobabilities (KDQ). Even in the simplest case, when the junction is initialized in a superposition of the ground and second excited states, the KDQ distribution of work exhibits negative or complex values, reflecting non-classical features. Moreover, the coherence content of the initial state can be optimized to enhance the extractable work extracted from the quench, beyond classical bounds. Finally, we propose an experimental interferometric protocol to directly measure the characteristic function of the work distribution in experimentally accessible settings. 
\end{abstract}

\maketitle

\section{Introduction}

The Josephson effect is one of the most striking examples of macroscopic coherence in quantum systems~\cite{josephson1962,feynman_lectures_JJ,BaroneBOOK}. In the realm of ultracold atoms, Bosonic Josephson Junctions (BJJs) can be realized by confining a Bose-Einstein condensate in a double-well potential~\cite{Zapata_Legget_1998,Gati2007,Tafuri2019}. Thanks to a high degree of controllability with long coherence times, BJJs have enabled the observation of a rich variety of fundamental phenomena~\cite{Trenkwalder2016,Spagnolli2017}, ranging from non-linear macroscopic quantum self-trapping~\cite{Smerzi1997} to atomtronic transport~\cite{CataliottiSCIENCE2001,Pezze2024}. Recently, the Josephson paradigm has also been extended to complex many-body phases, such as supersolid quantum gases featuring self-induced weak links~\cite{Biagioni2024,Donelli2025,Alana2025}. Beyond their equilibrium properties, BJJs can be an ideal testbed for out-of-equilibrium quantum dynamics~\cite{LenaPRA2016}. In the weakly interacting Rabi and Josephson regimes, the low-energy dynamics of BJJs can be accurately mapped, via the Holstein-Primakoff (HP) approximation, onto an effective time-dependent quantum harmonic oscillator (QHO)~\cite{OurReview}.

In \cite{DefenuGherardiniBuffoni2024,GomezRuiz2025}, it has been determined that such a mapping to a QHO can also be achieved for Lipkin-Meshkov-Glick critical systems, represented by fully-connected quantum spin models. For these systems, the work statistics originated by cyclic protocols crossing the quantum critical point are described by a negative binomial distribution, provided the system is initialized in the ground state of the undriven Hamiltonian. We therefore wonder whether such behaviour in the work statistics can also be observed in a driven BJJ, and we are interested in understanding how these statistics are affected by initialising the system in a quantum superposition state, before applying the work protocol.

In order to single-out the effects of quantum coherence in the system (initial) state, we consider Kirkwood-Dirac Quasiprobabilities (KDQs)~\cite{LevyPRXQuantum2020,LostaglioKirkwood2022,Gherardini2024Review,ArvidssonShukur2024review}, which provide results beyond the standard Two-Point Measurements (TPM) scheme~\cite{EspositoRMP2009,Campisi2011Review}. For non-autonomous quantum systems subjected to a coherent drive, the study of the KDQ distribution of the system internal energy allows one to identify contextual work instances, even leading to coherence-enhanced extractable work~\cite{SantiniPRB2023,HernandezPRR2024,DonatiSciRep2024,HernandezNpjQI2024,Pezzutto2025,GomezRuiz2025,SchemaInterferometrico,DonelliOrthoSpeedup}.

In this work, we study the quantum work statistics generated by a sudden quench transformation in BJJs encoded in a weakly interacting Bose-Einstein condensate. We describe the BJJ through the Bose-Hubbard model, and we map it to a time-dependent QHO using the HP approximation (Section \ref{sec:model}). Afterwards, we determine the analytical expression in closed-form of the work statistics (Section \ref{sec:quantum_work_statistic}). Initially, we assume that the BJJ is initialized in the ground state of the pre-quench Hamiltonian, demonstrating that the negative binomial distribution (recently identified in slow defect formation~\cite{DefenuGherardiniBuffoni2024}) also governs the work distribution originating from a sudden quench transformation. Then, to determine the role of quantum coherence in the system state before the quench is applied, we investigate the work statistics of the driven BJJ initialized in a superposition state of energy eigenstates. In such a condition, using the formalism of KDQs, we show that the presence of quantum coherence generally yields negative quasiprobabilities, possibly enhancing the amount of extractable work with respect to the incoherent case. We have benchmarked our theoretical predictions with the results obtained through the exact diagonalization of the two-mode Bose-Hubbard model, which also allowed us to identify the regime of validity for the harmonic HP approximation. Finally, in Section \ref{sec:exp_protocol}, we propose a feasible experimental protocol~\cite{Trenkwalder2016,Spagnolli2017,Macri2016} that is able to directly reconstruct the characteristic function of the work distribution, represented by a Loschmidt echo (LE) in our setting.

\section{Model}
\label{sec:model}

We consider a weakly interacting Bose-Einstein condensate confined in a symmetric double-well potential at low temperature. Assuming that the potential barrier is sufficiently high, the system can be described by the two-mode Bose-Hubbard Hamiltonian~\cite{Gati2007}
\begin{equation}\label{eq:BoseHubbard_Hamiltonian}
    \hat H = \frac{U}{4}\left(\hat n_l-\hat n_r\right)^2 -\frac{K}{2}\left(\hat c_l^\dagger \hat c_r + \hat c_r^\dagger \hat c_l\right),
\end{equation}
where $\hat c_{l}$ and $\hat{c}_{r}$ ($\hat{c}^\dagger_l$ and $\hat{c}^\dagger_r$) are the annihilation (creation) operators of the left and right wells, respectively. The corresponding number operators are $\hat n_{s}= \hat c_{s}^\dagger \hat c_{s}$, with $s=l,r$. In \eqref{eq:BoseHubbard_Hamiltonian}, $U$ denotes the self-interaction strength and $K$ is the tunnelling coefficient. We assume that the total number of atoms in the condensate is conserved (i.e., maintained constant) and equal to $N$. In the standard mean-field approach~\cite{Gati2007}, the operators $\hat{c}_{k}$ are replaced by complex numbers, resulting in a Hamiltonian that describes the physics of Josephson oscillations. For this reason, \eqref{eq:BoseHubbard_Hamiltonian} is also referred to as the BJJ Hamiltonian.

The two-mode Bose-Hubbard model with fixed number of particles is very well-suited to numerical studies. In fact, the associated Hilbert space has dimension $N+1$, where $N$ is the fixed number of atoms in the condensate. As a consequence, once an appropriate basis has been adopted, all the observables of the system are represented by Hermitian matrices of size $(N+1)\times(N+1)$, whose spectrum and eigenstates can be fully determined by exact diagonalization (see App.~\ref{App:Exact diagonalization}). The latter procedure will be performed numerically, and we will use it to benchmark our analytical predictions.

We now introduce the following effective spin operators:  
\begin{subequations}
\begin{align}
    \hat J_x = & \frac{1}{2}(\hat n_l-\hat n_r)\\
    \hat{J}_y = & \frac{i}{2}\left(\hat{c}_l^\dagger \hat{c}_r - \hat{c}_r^\dagger \hat{c}_l\right)\\
    \hat J_z = & \frac{1}{2}(\hat c_l^\dagger \hat c_r+\hat c_r^\dagger \hat c_l), 
\end{align}    
\end{subequations}
with the Casimir operator
\begin{equation}
    \hat{J}^2 = \hat{J}^2_x+\hat{J}^2_y+\hat{J}^2_z=\frac{N}{2}\left(\frac{N}{2}+1\right).
\end{equation}
These effective spin operators fulfil the standard angular momentum commutation relations $[\hat{J}_j,\hat{J}_k]=i\,\varepsilon_{jk\ell}\hat{J}_{\ell}$, where $j,k,\ell=x,y,z$. The Hamiltonian (\ref{eq:BoseHubbard_Hamiltonian}) can be rewritten in terms of $\hat J_x, \hat J_y, \hat J_z$ as 
\begin{equation}\label{eq:Spin_Hamiltonian}
    \hat H = U \hat J_x^2 - K \hat{J}_{z},
\end{equation}
which also arises in the context of quantum spin systems with all-to-all interactions, such as the Lipkin-Meshkov-Glick model~\cite{LIPKIN1965188,DusuelVidal2005}, and in atomic interferometry via the one-axis-twisting Hamiltonian~\cite{Kitagawa1993OAT}.

The ground state properties of \eqref{eq:BoseHubbard_Hamiltonian} strongly depend on the ratio between the self-interaction energy and the tunnelling parameter. Introducing the dimensionless parameter $\lambda \equiv NU/K$, we can distinguish three different regimes~\cite{Gati2007,LeggetBEC}:
\begin{itemize}
     \item $(1)$ Rabi regime: $\lambda\ll 1$
     \item $(2)$ Josephson regime: $1\ll\lambda\ll N^2$
     \item $(3)$ Fock regime: $\lambda\gg N^2$.
\end{itemize}
In the Rabi and Josephson regimes, it can be shown~\cite{G-SParaoanu_2001,DusuelVidal2005} that the ground state of \eqref{eq:Spin_Hamiltonian} is given by the $SU(2)$ coherent state of $\hat{J}_z$ with maximum value $N/2$. Hence, in these two regimes, we can apply an approximation around the semi-classical spin state, namely the Holstein-Primakoff approximation~\cite{HolsteinPrimakoff1940,IntroHP}. 

\subsection{Holstein-Primakoff approximation}

The HP transformation is a standard technique that enables one to describe the low-energy excitations of a spin system around a semi-classical spin configuration, e.g., a system fully polarized along the $z$-axis. This transformation makes use of a bosonic mode with annihilation and creation operators $\hat{a}$ and $\hat{a}^\dagger$, through which the angular momentum operators can be expressed as
\begin{subequations}
\begin{align}
    \hat{J}_{+} &= \sqrt{2j - \hat{a}^\dagger \hat{a}}\,\hat{a} \label{eq:exact_HP_transformation_1} \\
    \hat{J}_{-} &= \, \hat{a}^\dagger \sqrt{2j - \hat{a}^\dagger \hat{a}} \label{eq:exact_HP_transformation_2} \\
    \hat{J}_z &= (j - \hat{a}^\dagger \hat{a}) \,, \label{eq:exact_HP_transformation_3}
\end{align}    
\end{subequations}
where $j=N/2$, $\hat{J}_+=\hat{J}_x+i\hat{J}_y$ and $\hat{J}_-=\hat{J}^\dagger_+$. Under the assumption that the system remains close to the considered semi-classical spin configuration, the number of bosonic excitations is small compared to the total spin, such that $\langle a^\dagger a\rangle \ll 2j = N$. 
Using the approximation, we can perform a Taylor expansion of the square root operator and truncate it to the zeroth order, with the result that $\sqrt{1-\hat{a}^\dagger\hat{a}/(2j)} \simeq 1$. 
Substituting this expansion in the spin Hamiltonian of \eqref{eq:Spin_Hamiltonian}, the latter reduces to the Hamiltonian of a QHO:
\begin{equation}\label{eq:Hqho}
    \hat{H} \simeq \tilde{E} + \frac{1}{2m}\hat{p}^2 + \frac{1}{2}m\omega^2 \hat{q}^2,
\end{equation}
where we have introduced the dimensionless canonical conjugate variables $\hat{q} = \frac{1}{\sqrt{2}}(\hat{a} + \hat{a}^\dagger)$ and $\hat{p} = \frac{i}{\sqrt{2}}(\hat{a}^\dagger - \hat{a})$. Moreover, we have defined the effective mass $m = K^{-1}$, the plasma frequency $\omega = K\sqrt{ 1 + \lambda }$ and the energy offset $\tilde{E} = -\frac{K}{2}(N+1)$.

The mapping of the two-mode Bose-Hubbard Hamiltonian to the QHO is only approximate, as it is valid as long as $\langle \hat{a}^\dagger \hat{a} \rangle\ll N$, where the brackets denotes the average with respect to the ground state of \eqref{eq:Hqho}. Concerning the BJJ, it can be shown that the condition $\langle \hat{a}^\dagger \hat{a} \rangle\ll N$ is satisfied as long as the BJJ is in either the Rabi or Josephson regimes. 
Finally, the Hamiltonian (\ref{eq:Hqho}) expressed in terms of the canonical variables $\hat{q}$ and $\hat{p}$ offers some insights on the physical interpretation of the HP transformation. Considering the relation between $\hat{q}$ and $\hat{J_x}=\frac{1}{2}(\hat{n}_l-\hat{n}_r)$, we have
\begin{equation}\label{eq:relation_x_Jx}
    \hat{q}=\sqrt{\frac{2}{N}}\hat{J}_x=\frac{\hat{n}_l-\hat{n}_r}{\sqrt{2N}},
\end{equation}
i.e., the position operator $\hat{q}$ of the QHO is proportional to the atom-number difference between the two wells of the potential. As we will discuss in the following sections, the experimentally accessible observable $\hat{q}$ is directly related to thermodynamic quantities. 

\section{Quantum work statistics of quenched bosonic Josephson junctions}\label{sec:quantum_work_statistic}

Let us examine a sudden quench protocol for the Hamiltonian (\ref{eq:Hqho}), where the self-interaction $U$ is abruptly changed from an initial value $U_i$ to a final value $U_f$. The tunnelling constant $K$ is kept fixed, allowing us to disregard the constant energy shift $\tilde{E}$. 
We denote the pre-quench and the post-quench Hamiltonians as $\hat{H}_{i}$ and $\hat{H}_{f}$, which are given by
\begin{equation}\label{eq:pre-postQuenchHamiltonian}
    \hat{H}_{\gamma}=\frac{\hat{p}^2}{2m}+\frac{1}{2}m\omega^2_{\gamma} \hat{q}^2  \,\,\,\, \text{with } \,\,\,\, \gamma \in \{i,f\},
\end{equation}
where the mass $m$ and the plasma frequencies $\omega_i$ and $\omega_f$ are defined as in the previous section. It is a well-known result that the Hamiltonian of a QHO can be diagonalized by introducing the annihilation and creation operators 
\begin{equation}\label{eq:qhoannihilation/destruction}
    \hat{b}_\gamma=\sqrt{\frac{m\omega_\gamma}{2}}\hat{q} +i\frac{1}{\sqrt{2m\omega_\gamma}}\hat{p}
\end{equation}
that obey the canonical commutation relation $[\hat{b}_\gamma,\hat{b}^\dagger_\gamma] = 1$. The Hamiltonian $\hat{H}_{\gamma}$ can thus be rewritten as $\hat{H}_{\gamma} = \omega_\gamma (\hat{b}^\dagger_\gamma\hat{b}_{\gamma}+\frac{1}{2} )$, whose eigenstates $\ket{\Psi_n}$ (i.e., the Fock or number states) are
\begin{equation}
    \ket{\Psi^{(\gamma)}_n} = \frac{(\hat{b}^\dagger_\gamma)^n}{\sqrt{n!}}\ket{\Psi^{(\gamma)}_0},
\end{equation}
with corresponding eigenvalues $E^{(\gamma)}_n = \omega_\gamma\left(n+\frac{1}{2}\right)$. As a consequence, the Hamiltonian operators $\hat{H}_{i}, \hat{H}_{f}$ in \eqref{eq:pre-postQuenchHamiltonian} admit the spectral decomposition
\begin{equation}
    \hat{H}_{\gamma} = \sum_{n}^{}E^{(\gamma)}_n \ket{\Psi^{(\gamma)}_n}\!\bra{\Psi^{(\gamma)}_n}.
\end{equation}

We now assume that the system is initially prepared in a state described by a generic density operator $\hat{\rho}$, whose elements are given by $\rho_{j,\ell} \equiv \langle\Psi^{(i)}_j|\hat{\rho}|\Psi^{(i)}_\ell\rangle$ with respect to the pre-quench Hamiltonian eigenbasis. The initial state of the system may not commute with both the pre- and post-quench Hamiltonians. Accordingly, as also discussed in the Introduction, the statistics of work operated by the quench protocol has to be described by a probability distribution that goes beyond the TPM scheme. Here, we make use of KDQs that, in our analysis, are formally defined as
\begin{equation}
    q_{j,k} \equiv \braket{ \Psi^{(f)}_k }{\Psi^{(i)}_j} \bra{\Psi^{(i)}_j}\hat{\rho}\ket{\Psi^{(f)}_k}, 
\end{equation}
and are associated with the stochastic work $w_{j,k} \equiv E^{(f)}_k -E^{(i)}_j$. KDQs can be equivalently expressed in the more compact form 
\begin{equation}\label{eq:q_jk}
    q_{j,k} = \Lambda_{k,j}\sum_{\ell}^{}\Lambda^*_{k,\ell}\,\rho_{j,\ell} \,,
\end{equation}
where the overlaps (or transition amplitudes) between initial and final eigenstates of the pre- and post-quench Hamiltonians are denoted as $\Lambda_{m,n} = \langle\Psi^{(f)}_m|\Psi^{(i)}_n\rangle$.
Due to the possible non-commutativity of the system state with $\hat{H}_{\gamma}$ ($\gamma\in\{i,f\}$), the KDQs $q_{j,k}$ can assume non-positive (real negative or complex) values. As we will show below in Subsec.~\ref{sec:contextual}, there are circumstances in which non-positive values of $q_{j,k}$ can be regarded as non-classical, in reference to contextual work protocol allowing for coherence-enhanced extractable work~\cite{HernandezPRR2024}. In order to quantify the extent of non-positivity of KDQs $q_{j,k}$, we will consider the non-positivity functional~\cite{Gherardini2024Review} $\mathcal{N}\equiv -1 + \sum_{j,k}\left| q_{j,k} \right|$.

The KDQ distribution of work associated with the quench protocol reads as
\begin{equation}\label{eq:KDQworkDefinition}
   P\left[w\right] = \sum_{j,k}^{}q_{j,k}\,\delta\left(w-w_{j,k}\right),
\end{equation}
where $\delta(\cdot)$ is the Dirac delta function. Equivalently, the quantum work statistics can be described through the work characteristic function
\begin{equation}\label{eq:CFdefinition}
    \mathcal{G}_w(u) \equiv \sum_{j,k}^{}e^{iuw_{j,k}}q_{j,k}\,,
\end{equation}
which can be rewritten as a two-time correlation function~\cite{Campisi2011Review, Gherardini2024Review}:
\begin{equation}\label{eq:CharacteristicFunctionCorrelationFunction}
\mathcal{G}_w(u)=\operatorname{Tr}\left[e^{iu\hat{H}_{f}}e^{-iu\hat{H}_{i}}\hat{\rho}\right].
\end{equation}
It is worth noting that \eqref{eq:CharacteristicFunctionCorrelationFunction} is a LE evaluated over a generic density operator $\hat{\rho}$~\cite{LostaglioKirkwood2022,Gherardini2024Review,SchemaInterferometrico,DonelliOrthoSpeedup}.

In the following, we will study the quantum thermodynamics of the quenched BJJ considering different initial states: {\it (i)} the ground state of $\hat{H}_{i}$; {\it (ii)} a coherent superposition of $\hat{H_{i}}$'s energy eigenstates, which includes off-diagonal coherences in the energy basis of the system; {\it (iii)} the diagonal state obtained by removing the off-diagonal coherences from the superposition state at point (ii). The choice of these different classes of initial states serves to illustrate the role of quantum coherence in the non-equilibrium work statistics of quenched BJJs. 

\subsection{Overlaps between initial and final energy eigenstates}

Let us show how to analytically evaluate the overlaps $\Lambda_{m,n}$ required to compute the $q_{j,k}$ [\eqref{eq:q_jk}]. In doing this, we observe from \eqref{eq:qhoannihilation/destruction} that position and momentum operators, $\hat{q}, \hat{p}$, can be expressed in terms of the initial annihilation and creation operators $\hat{b}_\gamma,\hat{b}^\dagger_\gamma$. Then, knowing the decomposition of $\hat{q},\hat{p}$ in terms of the ladder operators $\hat{b}_i, \hat{b}_i^\dagger$ associated to the initial Hamiltonian $\hat{H}_{i}$, we are able to relate $\hat{b}_i, \hat{b}_i^\dagger$ with $\hat{b}_f, \hat{b}_f^\dagger$. The final result is:
\begin{equation}\label{eq:Bogoliubov}
\hat{b}_f = \cosh(r)\hat{b}_i+\sinh(r)\hat{b}^\dagger_i \,,
\end{equation}
where $r = \frac{1}{2}\ln( \omega_f / \omega_i)$; \eqref{eq:Bogoliubov} is a Bogoliubov canonical transformation. Using the Von Neumann theorem that every Bogoliubov canonical transformation can be represented by a suitable unitary operator~\cite{vonNeumann1931}, the transformation (\ref{eq:Bogoliubov}) can be represented in terms of the squeezing operator~\cite{OurReview}
\begin{equation}\label{eq:SqueezingOperatorDefinition.}
    \hat{S}(z)=\exp\left( -\frac{z}{2}\hat{b}_i^{\dagger2}+\frac{z^*}{2}\hat{b}_i^2 \right)
\end{equation}
with $z\in\mathrm{C}$. Since for quench protocols $z=r$ is a real number, the eigenstates of $\hat{H}_{f}$ are related to those of $\hat{H}_{i}$ by the relation $|\Psi^{(f)}_m\rangle = \hat{S}(r)|\Psi^{(i)}_m\rangle$. As a result, also the overlaps $\Lambda_{m,n}$ are real quantities that can be expressed as
\begin{equation}\label{eq:trans_amplitudes}
    \Lambda_{m,n}=\braket{\Psi^{(f)}_m}{\Psi^{(i)}_n}=\bra{\Psi^{(i)}_m}\hat{S}^{\dagger}(r)\ket{\Psi^{(i)}_n}.
\end{equation}
It can be observed that $\hat{S}^{\dagger}(r)=\hat{S}(-r)$; thus, the overlaps $\Lambda$ are equal to the matrix elements of a squeezing operator evaluated between number states. These matrix elements can be obtained analytically by using the disentangling theorem for the $SU(1,1)$ algebra, as shown in Refs.~\cite{Kim1989, Varro_2022}. Due to the parity of the number states, for the overlaps to be non-zero, $m-n$ must be an even number (i.e., $m-n=2k$ with $k \in \mathbb{Z}$). In the following, we will use the representation of $\Lambda_{m,n}$ in terms of associated Legendre polynomials $\mathcal{P}^{\abs{k}}_{l}(x)$~\cite{Varro_2022,OurReview}
\begin{equation}\label{eq:SqueezingFockLegendre}
    \Lambda_{m,n} = \text{sgn}\Big( (\omega_i-\omega_f)(m-n) \Big)^{\abs{k}} \sqrt{\frac{n_-!}{n_+!}} \sqrt{x}\,\mathcal{P}^{\abs{k}}_{l}(x), 
\end{equation}
where $k=\frac{m-n}{2}$, $l=\frac{m+n}{2}$, $x=\frac{1}{\cosh{(|r|)}}$, and we have defined $n_{-} \equiv \min(m,n)$, $n_{+} \equiv \max(m,n)$. 

\subsection{Initial ground state}\label{Subsectio:GroundState}

We now focus on the case where the BJJ is initialized in the ground state of $\hat{H}_{i}$, as in Ref.~\cite{LenaPRA2016}, such that the initial density operator can be written as $\hat{\rho}=|\Psi^{(i)}_0\rangle\!\langle\Psi^{(i)}_0|$. In this case, the KDQs reduce to the TPM joint probabilities
\begin{equation}\label{eq:KDQsAmplitudeGS}
    q_{0,k} = \left| \Lambda_{k,0} \right|^{2}.
\end{equation}
From the expression of the squeezing operator given in \eqref{eq:SqueezingOperatorDefinition.}, we can see that the overlap $\Lambda_{k,0}$ is zero when $k$ is odd. Hence, the explicit expression of the non-zero overlap for $k=2m$ (with $m$ positive integer) is 
\begin{equation}\label{eq:KDQsgroundstate}
    q_{0,2m} = \left|\Lambda_{2m,0}\right|^2 = \frac{(2m)!}{(m!\,2^m)^2}\frac{ \tanh(|r|)^{2m}}{\cosh(|r|)}.
\end{equation}
\eqref{eq:KDQsgroundstate} can be recognized as a negative binomial distribution with a fractional index~\cite{DefenuGherardiniBuffoni2024,GomezRuiz2025,OurReview}. To achieve this, we introduce the parameter 
\begin{equation}\label{eq:NBsuccesprob}
    \tilde{p}\equiv\frac{1}{\cosh{(|r|)}^2}
\end{equation}
by which $q_{0,2m}$ can be rewritten as
\begin{equation}\label{Eq:TPMNegativeBinomial}
    q_{0,2m} = \binom{m-\frac{1}{2}}{m} \tilde{p}^\frac{1}{2}(1-\tilde{p})^{m},
\end{equation}
which is the standard form of a negative binomial~\cite{degroot1986probability}.
The fractional index is a signature of the quantum nature of the process, since the negative binomial distribution with fractional index cannot be interpreted in the framework of Bernoulli trials~\footnote{In a classical scenario, $p,r \in \mathbb{N}$ and the negative binomial distribution $\text{prob}(k|r,p)=\binom{k+r-1}{k} p^r(1-p)^k$ gives the probability of having $k$ failures before exactly $r$ successes, in an infinite sequence of Bernoulli trials.}.

We can now write an analytical expression of the work probability distribution as
\begin{equation}\label{eq:NBWorkdistribution}
    P\left[w\right] = \sum_{m\in\mathrm{N}}^{}\binom{m-\frac{1}{2}}{m} \tilde{p}^\frac{1}{2}(1-\tilde{p})^m \delta(w-w_{0,2m}),
\end{equation}
which remains valid as long as the HP approximation holds. \eqref{eq:NBWorkdistribution} allows us to justify the approximation reported in \cite{LenaPRA2016} to describe the work distribution $P\left[w\right]$ for the BJJ initialized in the ground state of $\hat{H}_{i}$.

Having identified the work probability distribution as a negative binomial distribution, we can obtain a closed expression for the corresponding characteristic function.  
Using the definition of \eqref{eq:CFdefinition}, we have:
\begin{align}\label{eq:BJJcharacteristicfunction}
    \mathcal{G}_w(u) &= e^{iu\frac{\omega_f-\omega_i}{2}}\sum_{m}^{}e^{iu2m\omega_f}\binom{m-\frac{1}{2}}{m} \tilde{p}^\frac{1}{2}(1-\tilde{p})^m \nonumber \\
    &= e^{iu\frac{\omega_f-\omega_i}{2}}G_{\rm NB}(2\omega_fu),
\end{align}
where $G_{\rm NB}(t) \equiv \left( \frac{\tilde{p}}{1-(1-\tilde{p})e^{it}}\right)^{\frac{1}{2}}$ is the characteristic function of the negative binomial distribution~\cite{degroot1986probability}.
From Eqs.~(\ref{eq:NBWorkdistribution})-(\ref{eq:BJJcharacteristicfunction}), all the moments of the work distribution can be computed. In particular, the average work is 
\begin{equation}
    \left\langle w\right\rangle=\frac{m(\omega_f^2-\omega_i^2)}{2}\langle\hat{q}^2\rangle=\frac{\omega_f^2-\omega_i^2}{4\omega_i}=\frac{N}{4}\frac{\Delta U}{\omega_i}K,
\end{equation}
where we have used \eqref{eq:NBsuccesprob} and inserted the equality $r=\frac{1}{2}\ln{\left( \omega_f/\omega_i\right)}$. Moreover, the work variance reads as
\begin{eqnarray}
    \left( \Delta w\right)^2 &=& \frac{m^2(\omega_f^2-\omega_i^2)^2}{2}\langle \hat{q}^2\rangle^2=\frac{1}{8}\frac{\left(\omega_f^2-\omega_i^2\right)^2}{\omega_i^2} \nonumber\\
    &=&\frac{1}{8}\frac{N^2\Delta U^2}{\omega^2_i}K^{2}, 
\end{eqnarray}
in which we have substituted the explicit expression of the plasma frequency.

\begin{figure}[t]
    \centering
    \includegraphics[width=0.95\linewidth]{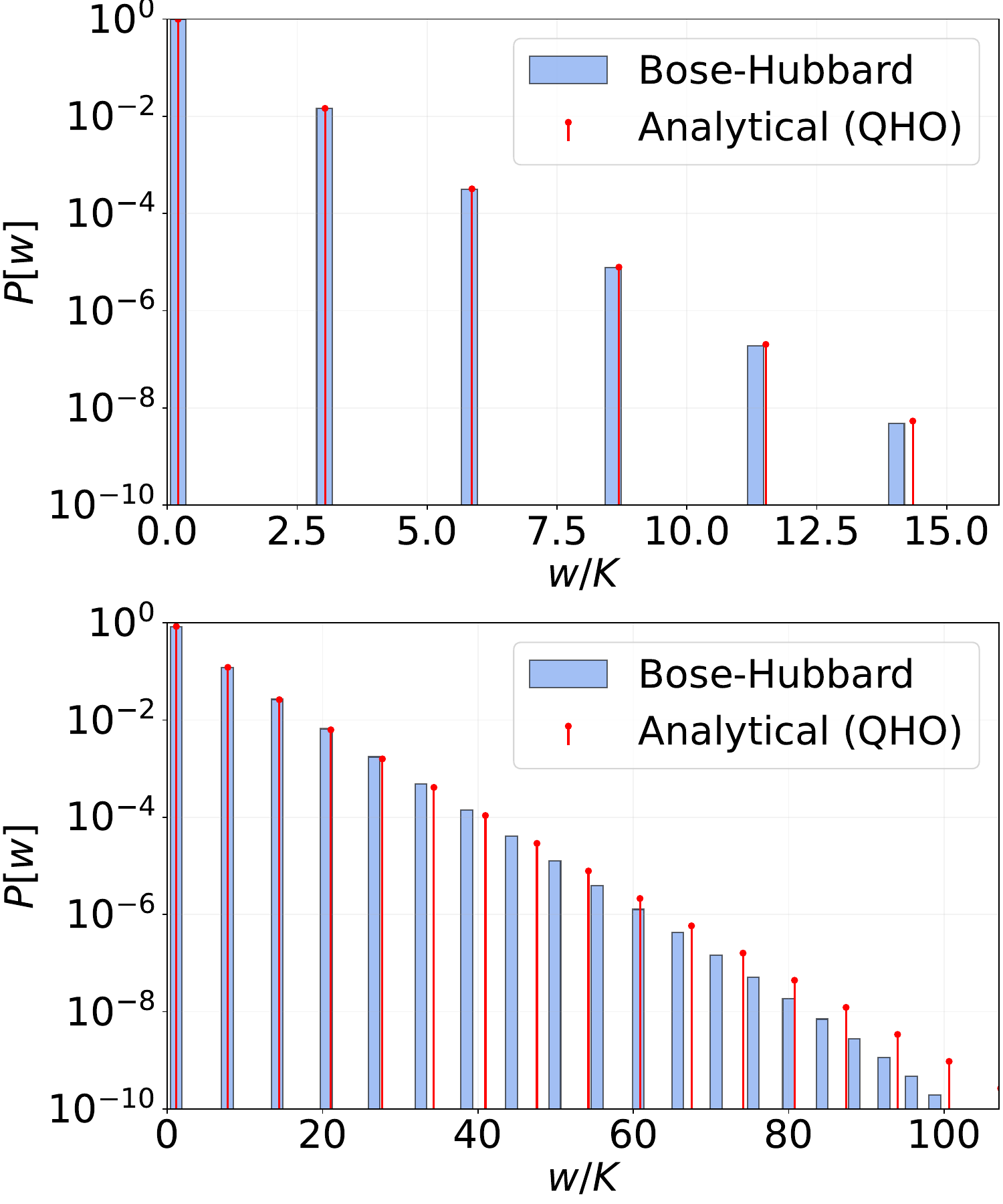}
    \caption{Work probability distributions for a sudden quench from the ground state. Comparison between the analytical (under the HP approximation) negative binomial distribution of \eqref{eq:NBWorkdistribution} (red stems) and the exact numerical distribution obtained via exact diagonalization of the Bose-Hubbard Hamiltonian in \eqref{eq:BoseHubbard_Hamiltonian} (blue bars). The parameter $U$ of the BJJ's Hamiltonian is quenched from an initial non-interacting state $U_i=0$ to $U_f=0.01 K$ (top panel) and $U_f=0.1K$ (bottom panel). The total number of particles is fixed at $N=100$.} 
    \label{fig:NB_quench}
\end{figure}

For a sudden quench protocol, the parameter $\tilde{p}$, which characterizes the negative binomial distribution and thus governs the work probability distribution, can be expressed as a function of the ratio of the final and initial plasma frequencies:
\begin{equation}
    \tilde{p}=\frac{4\frac{\omega_f}{\omega_i}}{\left(1+\frac{\omega_f}{\omega_i}\right)^2},
\end{equation}
with $\tilde{p}\in [0,1]$. In particular, $\tilde{p}\to 0$ for both $\frac{\omega_f}{\omega_i}\to 0$ and $\frac{\omega_f}{\omega_i}\to +\infty$. Hence, $\frac{\omega_f}{\omega_i}$ determines the spreading of the work probability distribution on its support. 
Moreover, $\frac{\omega_f}{\omega_i}$ effectively measures the strength of the quench: the larger the ratio, the stronger the quench. Increasing the quench strength favours the transition of the BJJ's state from the ground to highly excited states. As a consequence, our analytical predictions start to fail for very strong quenches, since the harmonic approximation captures only the low-energy excitations of the BJJ and cannot describe its full spectrum. 

In Fig.~\ref{fig:NB_quench}, we compare the analytical negative binomial distribution of work [\eqref{eq:NBWorkdistribution}] with the exact numerical results obtained through full diagonalization (see App.~\ref{App:Exact diagonalization}) of the Bose-Hubbard Hamiltonian (\ref{eq:BoseHubbard_Hamiltonian}). We analyse two work protocol given by the quench of $U$ with different strengths, fixing the initial self-interaction parameter to $U_i=0$ (this means that the BJJ is initially in the Rabi regime). The tail of the corresponding work distributions corresponds to the transition from the ground state of $\hat{H}_{i}$ to highly excited states of $\hat{H}_{f}$. 

In the top panel of Fig.~\ref{fig:NB_quench}, the final self-interaction parameter $U_f$ is such that the system is still in the Rabi regime, while in the bottom panel the BJJ is driven into the Josephson regime. The analytical work probability distribution (\ref{eq:NBWorkdistribution}) is in good agreement with the numerics for the intra-regime quench (Rabi to Rabi). In such a case, the low-energy spectrum of \eqref{eq:BoseHubbard_Hamiltonian} is effectively linear and the energy difference between consecutive eigenvalues is well approximated by the plasma frequency~\cite{Gati2007}, making the HP approximation very accurate. 

On the other hand, applying a quench transformation from the Rabi to the Josephson regime, we observe a progressive shift on the right of the stochastic work values determined analytically using the HP approximation, with respect to the exact ones obtained through diagonalization. This shift becomes larger as $w$ increases, and is due to the fact that the exact many-body energy gaps deviate from the uniform harmonic spacing of the QHO. These discrepancies arise from the breakdown of the zeroth-order HP approximation, which cannot describe the high energy ($E > NK$) spectrum and eigenstates of the BJJ Hamiltonian, \eqref{eq:BoseHubbard_Hamiltonian}. 

\subsection{Initial superposition state}

We assume that the system is initialized in a pure state, superposition of  two eigenstates of $\hat{H}_{i}$. The selection rules on the overlaps, \eqref{eq:trans_amplitudes}, require the two components of the initial superposition to have the same parity, such that the interference terms in the KDQs are not cancelled. Otherwise, the standard TPM scheme for the characterization of the work statistics suffices. Moreover, as the HP approximation is able to well-describe the low-energy excitations of the Bose-Hubbard Hamiltonian, we have selected as initial state the superposition of the ground and second excited states. Hence, we initialize the BJJ in the state
\begin{equation}\label{eq:SuperposTsate02}
    \ket{\Psi_0}=\alpha \ket{\Psi^{(i)}_0}+\beta\ket{\Psi^{(i)}_2}.
\end{equation}
Notice also that, being quite challenging to create high energy states in experimental settings, the superposition state (\ref{eq:SuperposTsate02}) is also a viable experimental option.

The initial state (\ref{eq:SuperposTsate02}) guarantees that the density operator $\hat{\rho}=|\Psi_0\rangle\!\langle\Psi_0|$ does not commute with the initial Hamiltonian $\hat{H}_{i}$, thereby allowing interference effects induced by quantum coherence to appear in the work statistics. The KDQs, \eqref{eq:q_jk}, describing the work probability distribution of a quenched BJJ initialized in the initial superposition state (\ref{eq:SuperposTsate02}) read as
\begin{subequations}\label{eq:KDQslinearsuperposition}
\begin{align}
    & q_{0,2m}=\left|\alpha\right|^2 \left|\Lambda_{2m,0}\right|^2 +\alpha\beta^*\Lambda_{2m,0}\Lambda_{2m,2}\\
    & q_{2,2m}=\alpha^*\beta\,\Lambda_{2m,0}\Lambda_{2m,2} + \left|\beta\right|^2 \left|\Lambda_{2m,2}\right|^2 \,.
\end{align}    
\end{subequations}
In general, $q_{0,2m}$ and $q_{2,2m}$ are complex numbers, being $\alpha,\beta \in \mathbb{C}$. An analytical expression for the ratio between the overlaps $\Lambda_{2m,2}$ and $\Lambda_{2m,0}$ can be obtained (see App.~\ref{APP:TA_ratio}): 
\begin{equation}\label{eq:AmplitudesRatio}
  \frac{\Lambda_{2m,2}}{\Lambda_{2m,0}}=g_m(x)=\text{sgn}\left( \omega_i-\omega_f\right)\frac{1-x^2(1+2m)}{\sqrt{2(1-x^2)}},
\end{equation} 
where, we recall, $x=\frac{1}{\cosh{(|r|)}}$. Using \eqref{eq:AmplitudesRatio}, the KDQs (\ref{eq:KDQslinearsuperposition}) simplify to
\begin{subequations}\label{eq:KDSuperpos02vers1}
\begin{align}
    q_{0,2m} =& \left|\Lambda_{2m,0}\right|^2 \Big[ |\alpha|^2 + g_m(x)\,\alpha\beta^* \Big]\\ 
    q_{2,2m}  =& \left|\Lambda_{2m,0}\right|^2 g_m(x) \Big[ g_m(x) \,|\beta|^2 + \alpha^*\beta \Big].
\end{align}    
\end{subequations}
Interestingly, using \eqref{Eq:TPMNegativeBinomial}, the KDQs (\ref{eq:KDSuperpos02vers1}) can be interpreted as weighted combinations of the negative binomial distribution $|\Lambda_{2m,0}|^2$.

Having characterized the KDQs, let us analyse the average work and the work variance of the KDQ distribution of work. The average work is given by~\cite{Gherardini2024Review}
\begin{equation}\label{eq:AverageWork_KD}
    \langle w\rangle=\sum_{j,k}^{}q_{j,k}\, w_{j,k}=\operatorname{Tr}\left[\left(\hat{H}_{f}-\hat{H}_{i}\right)\hat{\rho}\right],
\end{equation}
which is valid for a generic initial density matrix $\hat{\rho}$ (in our case, built over a pure state). It is known that, even if $q_{j,k}$ may be complex numbers, the average work is always real.  
For the specific superposition state in \eqref{eq:SuperposTsate02}, the average work splits into two contributions: i) a term indicated as $\langle w \rangle_{\text{D}}$ arising from the diagonal part (population terms) $\hat{\rho}_D = |\alpha|^2 \, |\Psi^{(i)}_0\rangle\!\langle\Psi^{(i)}_0| + |\beta|^2 \, |\Psi^{(i)}_2\rangle\!\langle\Psi^{(i)}_2|$ of the initial density operator, and ii) a purely coherent term, denoted by $\langle w\rangle_{\text{C}}$, that takes into account the off-diagonal coherence terms of $\hat{\rho}$. 
Explicitly, the average work is given by $\langle w\rangle=\langle w\rangle_{D}\,+\langle w\rangle_C$ where
\begin{subequations}
\begin{align}
    \langle w \rangle_{D} &=\frac{\omega_f^2-\omega_i^2}{2\omega_i}\left( 2|\beta|^2+\frac{1}{2}\right) = \frac{N}{2}\frac{\Delta U}{\omega_i}K\left( 2|\beta|^2 + \frac{1}{2} \right)\\
    \langle w \rangle_{C} &=\frac{N\sqrt{2}}{2}\frac{\Delta U}{\omega_i}K\Re\left[ \alpha\beta^* \right].
\end{align} 
\end{subequations}
The diagonal contribution $\langle w \rangle_{D}$ can be also determined through the TPM scheme, which however cannot access the coherent contribution $\langle w \rangle_{C}$ that is proportional to $\Re[\alpha\beta^*]$. The latter fact means that the average work depends also on the relative phase between the energy eigenstate of the initial superposition state. Consequently, quantum coherence can either enhance or suppress the average work.

The evaluation of the work variance is more involved, since it requires the computation of the second statistical moment $\langle w^2 \rangle=\sum_{j,k}^{}q_{j,k} w^2_{j,k}$. As shown in \cite{Gherardini2024Review}, $\langle w^2 \rangle$ contains a two-time correlation function and can become complex. Moreover, the real part of $\langle w^2 \rangle$ can be written in the following physically-intuitive way:
\begin{equation}\label{eq:secondmomentRealPart}
    \Re[\langle w^2 \rangle]=\operatorname{Tr}\left[\left(\hat{H}_{f}-\hat{H}_{i}\right)^2\hat{\rho}\right].
\end{equation}
For the state in \eqref{eq:SuperposTsate02}, the second moment of work again splits into a diagonal (or TPM) contribution and a coherence-induced one, i.e., $\Re[\langle w^2\rangle]=\langle w^2\rangle_{D}+\langle w^2\rangle_C$ where 
\begin{subequations}
\begin{align}
  \langle w^2\rangle_{D} = &\frac{3}{16}\frac{N^2\Delta U^2}{\omega_i^2}K^2\left( 1+12|\beta|^2 \right)\\
    \langle w^2\rangle_{C} = &\frac{3\sqrt{2}}{4}\frac{N^2\Delta U^2}{\omega_i^2}K^2\Re\left[ \alpha\beta^* \right].
\end{align} 
\end{subequations}
As for the average work, $\Re[\langle w^2\rangle]$ contains a contribution proportional to $\Re[\alpha\beta^*]$, entailing that quantum coherence entering the initial state affects both the average work and work fluctuations. Finally, the real part of the work variance $(\Delta w)^2$ is obtained as
\begin{equation}
    \Re[\left(\Delta w\right)^2]=\Re[\langle w^2 \rangle]-\langle w \rangle^{2}.
\end{equation}

In order to show how quantum coherence in the initial state (with respect to the basis of $\hat{H}_{i}$) modifies the work statistics, we consider a quench protocol from the Josephson ($U_i=0.1K$) to the Rabi regime ($U_f=0K$) of the BJJ, as done in Subsec.~\ref{Subsectio:GroundState}.
Notice that the average work done by the system is $\mathcal{W} = -\langle w \rangle$ such that, by convention, when $\mathcal{W}>0$, work is extracted on average from the system. In App.~\ref{App: Optimization} it is shown that for the quench transformations considered here, the average work done by the system is positive for all the states of type \eqref{eq:SuperposTsate02}, meaning that work is always extracted. 

\begin{figure*}[t]
    \includegraphics[width=0.495\linewidth]{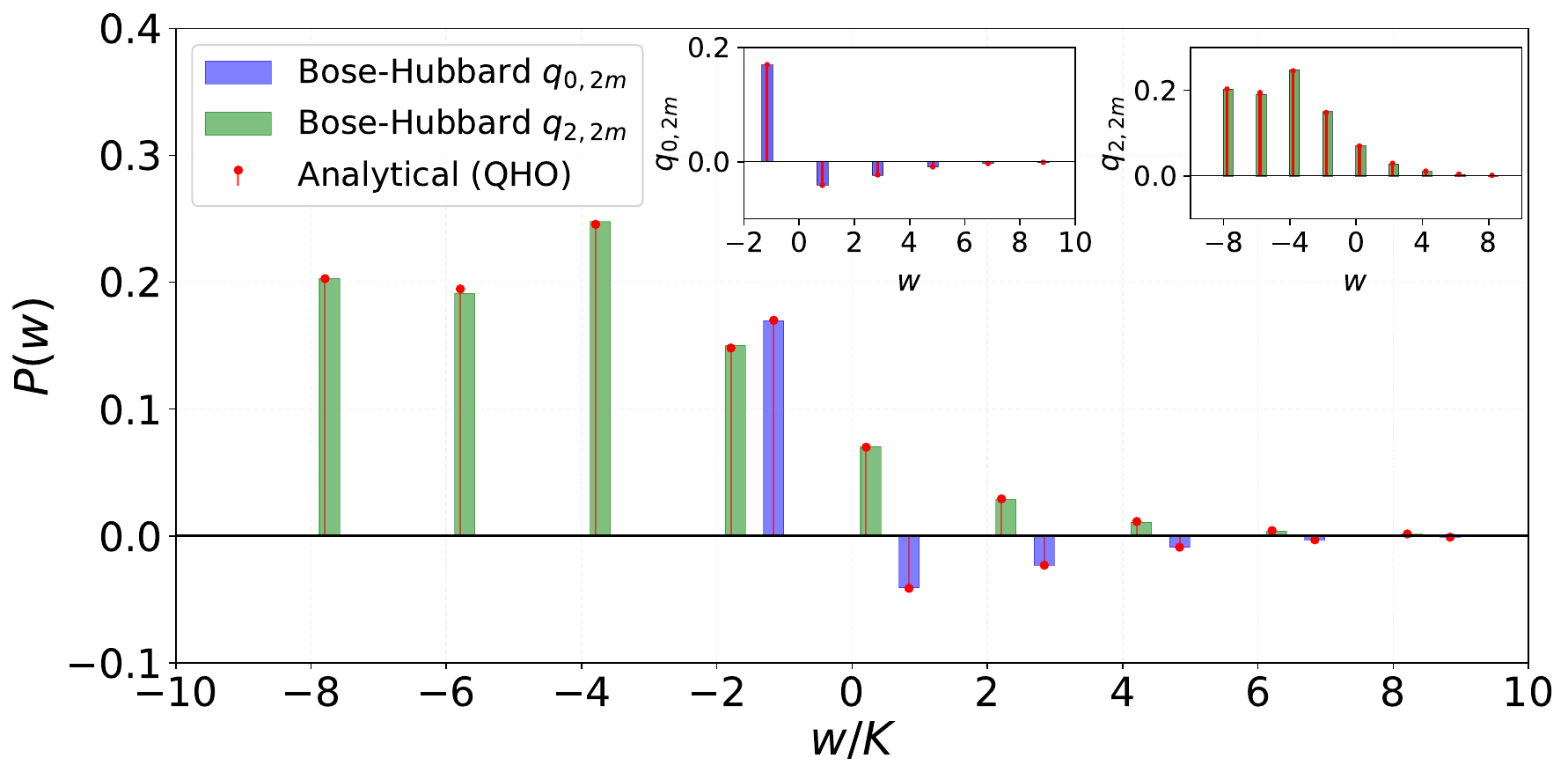}
    \includegraphics[width=0.495\linewidth]{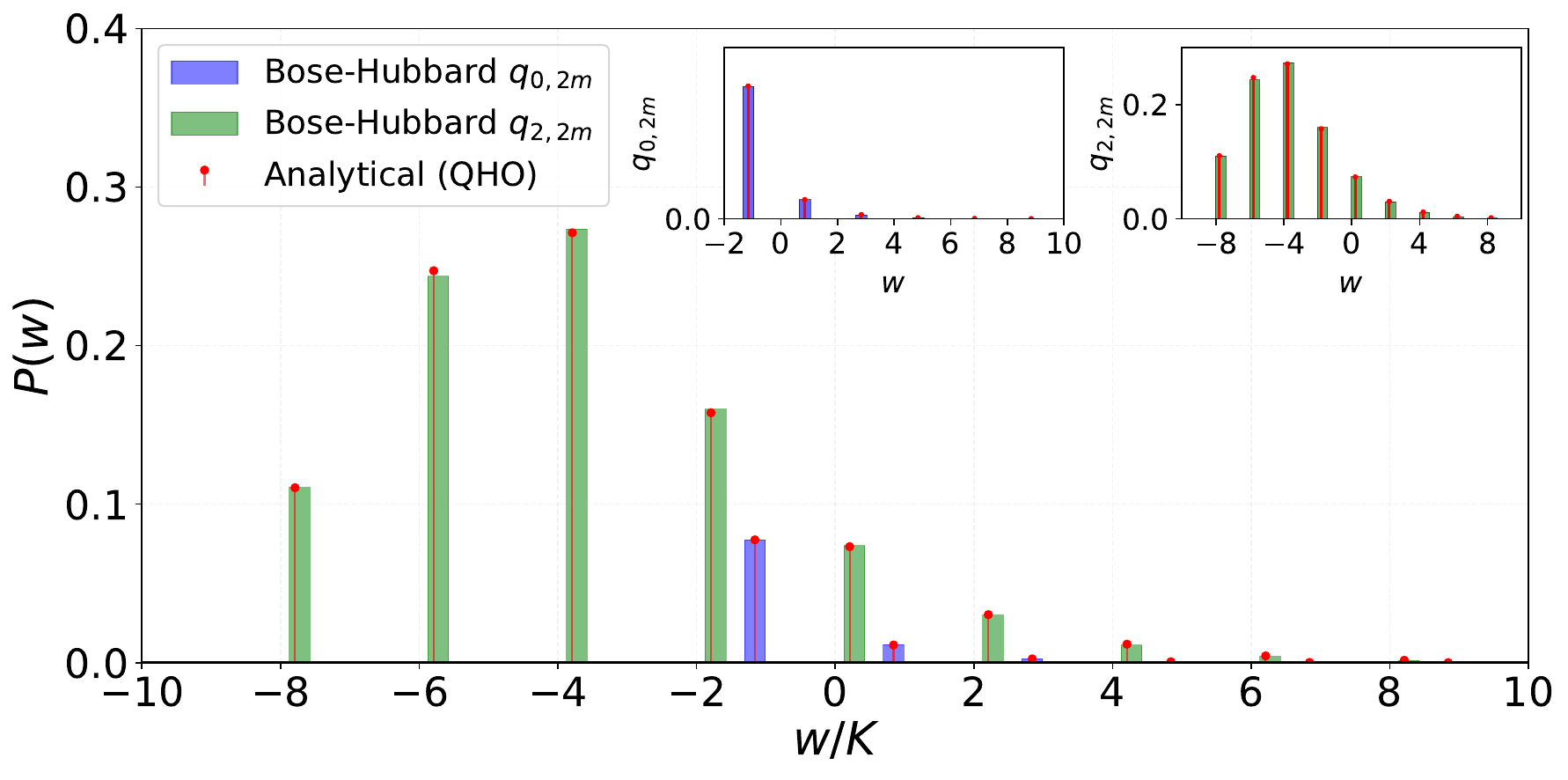}
    \caption{KD quasiprobability distributions of work, \eqref{eq:KDQworkDefinition}, considering the superposition state (\ref{eq:SuperposTsate02}) (top panel) or the diagonal mixed state with same population terms (bottom panel) as the initial state. 
    The parameters of the quench protocol are $U_i=0.1K$ (Josephson regime), $U_f=0K$ (Rabi regime), and $N=100$. The coloured bars are the KDQs obtained from the exact diagonalization of the Bose-Hubbard Hamiltonian (\ref{eq:BoseHubbard_Hamiltonian}). The red stems are the analytical expression of the KDQs, as given in \eqref{eq:KDSuperpos02vers1}. In the inset of both panels, we plot the single KDQ $q_{0,2m}$ (blue bars) and $q_{2,2m}$ (green bars) for the non-zero stochastic work instances.}
    \label{fig:KDworkdistributions}
\end{figure*}

In App.~\ref{App: Optimization}, $\mathcal{W}$ is maximized with respect to the coefficients $\alpha, \beta$ of the superposition state
(\ref{eq:SuperposTsate02}), obtaining $\alpha_{\rm opt} = 0.316$ and $\beta_{\rm opt} = 0.949$. Interestingly, the KDQs associated to this superposition state are real, since the overlaps $\Lambda$ are real quantities and $\alpha_{\rm opt},\beta_{\rm opt} \in \mathbb{R}$. 
In order to understand the interference effects due to quantum coherences in the initial state, we compare in Fig.~\ref{fig:KDworkdistributions} the work statistics obtained initializing the BJJ in the superposition state (\ref{eq:SuperposTsate02}) with $\alpha_{\rm opt},\beta_{\rm opt}$, or in the diagonal mixed state $\hat{\rho}_D = |\alpha_{\rm opt}|^{2} |\Psi^{(i)}_0\rangle\!\langle\Psi^{(i)}_0| + |\beta_{\rm opt}|^{2}|\Psi^{(i)}_2\rangle\!\langle\Psi^{(i)}_2|$.

In Fig.~\ref{fig:KDworkdistributions}, we can observe that $\mathcal{W}$ is larger for the case of an initial superposition state ($\mathcal{W}=4.05K$) with respect to initializing the BJJ in the corresponding diagonal mixed state ($\mathcal{W}=3.43K$). This entails a boost in the extractable work arising from quantum interference effects.
In Ref.~\cite{Gherardini2024Review}, it has been shown that the real part of the work variance can exhibit local minima when there are quantum coherences in the system state decomposed in the $\hat{H}_{i}$ basis. 
We observe this behaviour also when applying a sudden quench transformation on a BJJ, as we have considered here: the real part of the variance $\Re[\left(\Delta \mathcal{W}\right)^2] = 6.06K^2$ is smaller for the initial superposition state than for the initial diagonal state whereby $\Re[\left(\Delta \mathcal{W}\right)^2] = 7.90K^2$.

Looking at the whole distribution in Fig.~\ref{fig:KDworkdistributions}(a), we can notice that the KDQs assume negative values, signature of non-classical work probability distribution. 
The presence of non-positive values of the work distribution can be quantified by the non-positivity functional $\mathcal{N}$, defined in Sec.~\ref{sec:quantum_work_statistic}. In particular, for the setting analysed in Fig.~\ref{fig:KDworkdistributions}(a), we find $\mathcal{N}=0.156$. 
These features are absent in Fig.~\ref{fig:KDworkdistributions}(b), where we plot the probability distribution of work corresponding to the initial diagonal mixed state $\hat{\rho}_D$.

As shown in \cite{Arvidsson-Shukur_2021,Gherardini2024Review}, the presence of negative values in the KDQs can be traced back to the non-commutativity between $\hat{\rho}=|\Psi_0\rangle\!\langle\Psi_0|$ and $\hat{H}_{i}$, i.e., $[\hat{\rho},\hat{H}_{i}]\neq 0$ is a necessary condition for the loss of positivity in the work distribution. Finally, from Fig.~\ref{fig:KDworkdistributions}(a), it is worth observing that the effect of maximizing $\mathcal{W}$ is that all the negative KDQs are associated to positive values of work. This means that an excitation process occurring with negative $q_{j,k}$ is equivalent to a de-excitation process in a classical work protocol with probability $|q_{j,k}|$. Since any de-excitation process increases the amount of extractable work $\mathcal{W}>0$, the presence of negativity leads to an extractable work enhancement. 

\subsection{Contextuality and coherence-enhanced extractable work}
\label{sec:contextual}

The presence of negative real or complex values of the KDQ work distribution can lead to an enhancement of extractable work $\mathcal{W}>0$~\cite{SantiniPRB2023,HernandezPRR2024,DonatiSciRep2024,Gherardini2024Review,HernandezNpjQI2024,Pezzutto2025,GomezRuiz2025,SchemaInterferometrico,DonelliOrthoSpeedup}. Moreover, as shown in \cite{HernandezPRR2024}, such enhancement can be witnessed by the violation of a bound on the maximum extractable work obtained under the conditions $[\hat{\rho},\hat{H}_{i}]=0$ and $[\hat{H}_{i},\hat{H}_{f}]=0$. This bound can be saturated, but not overcome, by work protocols implementing classical probabilistic operations that are associated to real joint probabilities $\in [0,1]$ of the work distribution. As a consequence, when the non-positivity function is identically equal to zero, $\mathcal{N}=0$, the following relation holds:
\begin{equation}\label{eq:UpperBoundWorkExtraction}
    \mathcal{W} \le \mathcal{W}^{\rm max}_{\rm clas} = - \sum_{i,j|w_{j,k}<0} w_{j,k}\sqrt{p_{j,k}\,p_k}
\end{equation}
where, in our setting, $p_{j,k} = |\Lambda_{k,j}|^2 \, |\langle\Psi^{(i)}_j|\Psi_0\rangle|^2$ are the joint probabilities of the TPM work distribution, and $p_k = |\langle\Psi^{(f)}_k|\Psi_0\rangle|^2$.

For the initial superposition state of \eqref{eq:SuperposTsate02}, the bound (\ref{eq:UpperBoundWorkExtraction}) admits a closed-expression with a clear physical interpretation. In fact, using the selection rules on the overlaps $\Lambda_{k,j}$, we must have $k=2m$ with $m$ a positive integer, such that from \eqref{eq:AmplitudesRatio} we get:
\begin{subequations}\label{eq:TPMprobbound}
\begin{align}
    p_{0,2m} =& \left|\Lambda_{2m,0}\right|^2 |\alpha|^2 \\
    p_{2,2m} =& g^2_m(x)\left|\Lambda_{2m,0}\right|^2|\beta|^2\\
    p_{2m} =& f^2_m\left(x,\alpha,\beta\right)\left|\Lambda_{2m,0}\right|^{2},
\end{align}    
\end{subequations}
with $f_m(x,\alpha,\beta) \equiv \sqrt{|\alpha|^2+g_m^2(x)|\beta|^2+2g_m(x)\Re[\alpha\beta^*]}$, and we recall $x=\frac{1}{\cosh(|r|)}$ with $r=\frac{1}{2}\ln(\omega_f/\omega_i)$.
In this way, \eqref{eq:UpperBoundWorkExtraction} reads as 
\begin{eqnarray}\label{eq:explicitBound}
\mathcal{W}_{\rm clas}^{\rm max} = -\sum_{m=0}^{}\Theta (-w_{0,2m})w_{0,2m}\left|f_m\left(x,\alpha,\beta\right)\right|\left|\Lambda_{2m,0}\right|^2 \left|\alpha\right|\nonumber\\
-\sum_{m=0}^{}\Theta (-w_{2,2m})w_{2,2m}\left|g_m(x)f_m(x,\alpha,\beta)\right|\left|\Lambda_{2m,0}\right|^2\left|\beta\right|,\quad
\end{eqnarray}
where $\Theta(\cdot)$ denotes the Heaviside theta function. The bound $\mathcal{W}^{\rm max}_{\rm clas}$ of \eqref{eq:explicitBound} is model-dependent and depends on the ratio of plasma frequencies associated to the initial and final Hamiltonians. It is also a function of the coefficients $\alpha,\beta$ of the superposition state (\ref{eq:SuperposTsate02}). The expression of \eqref{eq:explicitBound} suggests a possible expression for the positive joint probabilities, which we denote as $p^{\rm clas}_{j,k}$, that saturates the bound:
\begin{equation}\label{eq: explicitJointProb}
    \begin{split}
        p^{\rm clas}_{0,2m} =& \Theta(-w_{0,2m})\left|f_m(x,\alpha,\beta)\right| \left|\Lambda_{2m,0}\right|^2 \left|\alpha\right|\\
        p^{\rm clas}_{2,2m} =& \Theta(-w_{2,2m})\left|g_m(x)f_m(x,\alpha,\beta)\right| \left|\Lambda_{2m,0}\right|^2 \left|\beta\right|,
    \end{split}
\end{equation}
and all the other $p^{\rm clas}_{j,k}=0$, such that the bound  (\ref{eq:explicitBound}) can be formally expressed as 
\begin{equation}\label{eq:bound_TPM}
    \mathcal{W}^{\rm max}_{\rm clas}=-\sum_{j,k}^{}p^{\rm clas}_{j,k}\,w_{j,k}\,.
\end{equation}
\eqref{eq:bound_TPM} makes explicit the classical nature of $\mathcal{W}^{max}_{\rm clas}$. It is worth noting that $p^{\rm clas}_{j,k}$ are different from the joint probabilities $p_{j,k}$ in \eqref{eq:TPMprobbound}, which are the KDQs associated to the diagonal mixed state $\hat{\rho}_D$. Thus, $p_{j,k}$ can be obtained from initializing the system in the density operator $\hat{\rho} = |\Psi_0\rangle\!\langle\Psi_0|$ and then implementing the TPM scheme~\cite{Gherardini2024Review}.

The violation of \eqref{eq:UpperBoundWorkExtraction} represents a proof of the contextuality of quantum theory~\cite{Spekkens2005}, in relation to the extractable work generated by a quench protocol applied to a BJJ. Such a violation is witnessed by the loss of positivity of the corresponding KDQ distribution of work, and thus by the presence of negative KDQs resulting in $\mathcal{N} \ge 0$. 

\begin{figure}[t]
    \centering
    \includegraphics[width=0.95\linewidth]{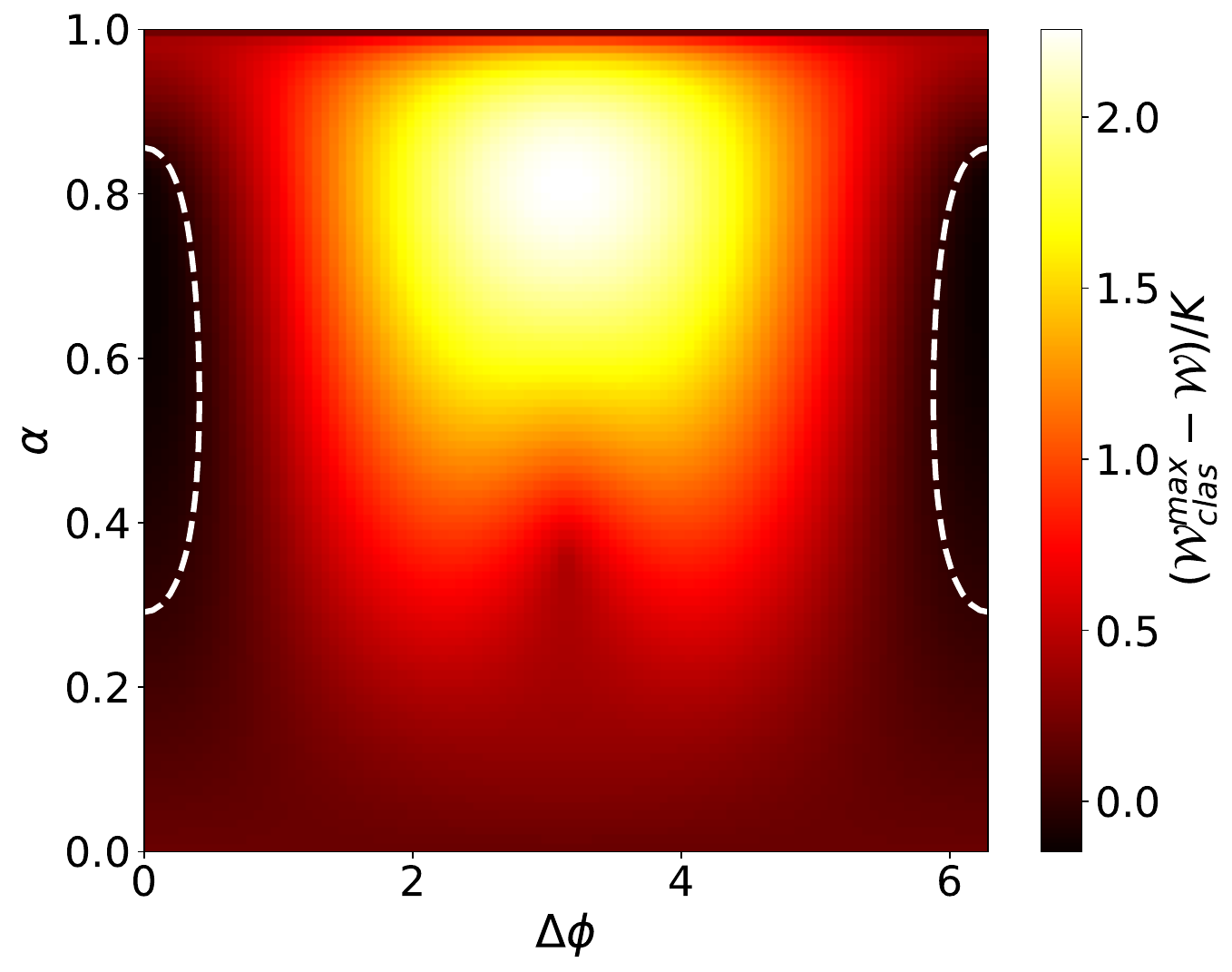}
    \caption{Surface plot of $\mathcal{W}^{\rm max}_{\rm clas}-\mathcal{W}$ as a function of $\alpha$ (coefficient of the initial superposition state $|\Psi_0\rangle$, \eqref{eq:SuperposTsate02}) and of the phase difference $\Delta\phi$ between the ground and the second excited states entering $|\Psi_0\rangle$. The dashed white lines mark the boundaries of the region where $\mathcal{W}^{\rm max}_{\rm clas}-\mathcal{W}<0$, signalling a violation of the bound (\ref{eq:explicitBound}). The analytical predictions are in very-good agreement with the results obtained through the diagonalization procedure.}
    \label{fig:ContextualityViolation}
\end{figure}

In order to illustrate this finding, let us consider again the quench protocol with $U_i = 0.1K$, $U_f=0$ and $N=100$ atoms. Using the optimization described in App.~\ref{App: Optimization}), we can ensure that $\mathcal{W}>0$. We perform exact diagonalization to \eqref{eq:BoseHubbard_Hamiltonian}, and we evaluate the quantity $\mathcal{W}^{\rm max}_{\rm clas}-\mathcal{W}$ by varying $\alpha \in [0,1)$ and $\Delta\phi\in [0,2\pi)$, where $\Delta\phi$ is the phase difference between the coefficients of the ground and second excited state in the initial superposition state $|\Psi_0\rangle$. If $\mathcal{W}^{\rm max}_{\rm clas} - \mathcal{W}<0$ for some values of $\alpha$ and $\phi$, then the KDQs associated to the initial state (here, a superposition state) are non-positive, and the work protocol that gives rise to $\mathcal{W}$ is contextual in the sense of Ref.~\cite{Spekkens2005}.
As one observes from Fig.~\ref{fig:ContextualityViolation}, we can identify regions in the $\Delta\phi-\alpha$ plane where the extractable-work bound (\ref{eq:explicitBound}) is violated. The maximum violation is achieved for $\alpha=0.697$ and $\Delta\phi=0$. Notice that the quantum state for which the work protocol is maximally contextual does not necessarily correspond to the state that maximizes the amount of extractable work (see also App.~\ref{App: Optimization}). Interestingly, in our example, the state that maximizes the extractable work does not violate \eqref{eq:UpperBoundWorkExtraction}, meaning that the maximum amount of extractable work could be obtained in principle through a transformation with a classical stochastic description.

\section{Experimental protocol proposal}
\label{sec:exp_protocol}

\begin{figure}[t]
    \centering
    \includegraphics[width=1.01\linewidth]{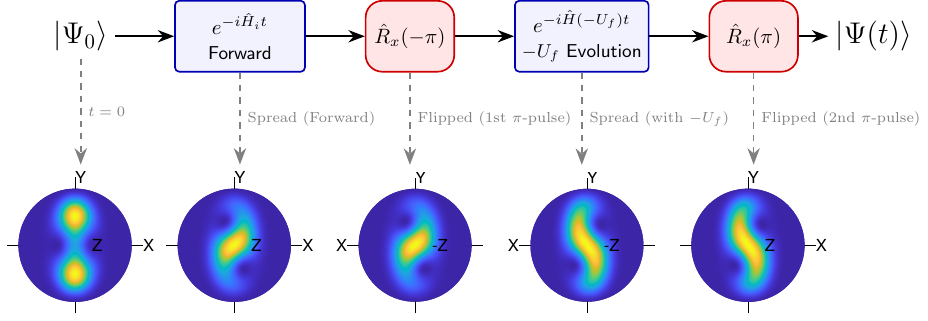}
    \includegraphics[width=\linewidth]{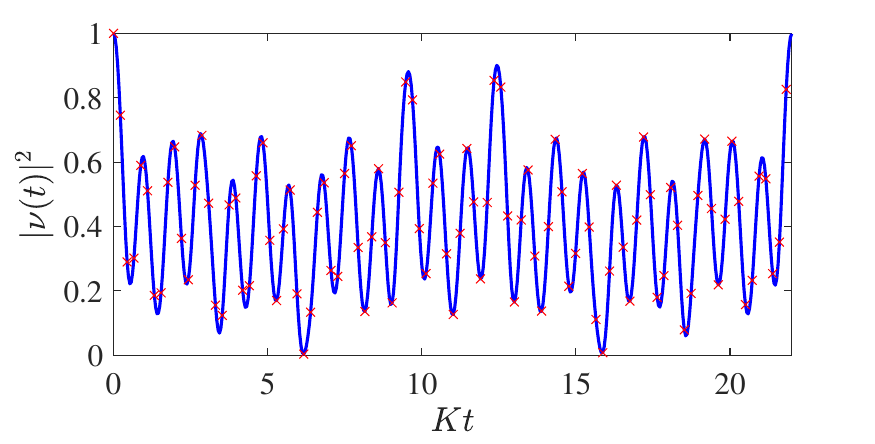}
    \caption{Proposed experimental protocol for the direct measurement of the LE. \textit{(Top)} Block diagram representing the time evolution sequence. The backward time evolution $e^{i\hat{H}_f t}$ is achieved by inverting the interaction parameter, $U_f \rightarrow -U_f$, and enclosing the system evolution between two $\pi$-pulses around the $\hat{J}_x$ axis. \textit{(Middle)} Husimi phase-space distributions mapped on the Bloch sphere at different stages of the experimental protocol. They illustrate the phase-space stretching during the forward evolution, the flipping via the $\pi$-pulses and the subsequent partial refocusing. \textit{(Bottom)} Corresponding LE amplitude $|\nu(t)|^2$ as a function of the dimensionless time $Kt$. The solid blue line represents the exact numerical simulation of the ideal evolution $e^{i\hat{H}_f t}e^{-i\hat{H}_i t}$ in \eqref{eq:LE}, while the red crosses represent the results from numerically implementing the proposed experimental sequence. The perfect agreement validates the experimental feasibility of our proposal.}
    \label{fig:ExpProtocol_Full}
\end{figure}

To experimentally reconstruct the characteristic function of the work distribution, we propose a protocol able to directly measure the LE [\eqref{eq:CharacteristicFunctionCorrelationFunction}] associated with the quench transformation. The complete sequence of the experimental protocol is illustrated in Fig.~\ref{fig:ExpProtocol_Full}. The top panel of Fig.~\ref{fig:ExpProtocol_Full} shows the experimental operations, the middle panel provides a phase-space representation of the dynamics through Husimi distributions on the Bloch sphere, while the bottom panel benchmarks the LE obtained from the proposed experimental sequence against the numerically simulated one by direct calculation. 

Let us recall the Hamiltonian of the BJJ written in terms of spin operators: $\hat{H}_{i(f)} = U_{i(f)} \hat{J}_x^2 - K \hat{J}_z$, representing the system before (after) the quench transformation. For a quench from $U_i$ to $U_f$, the LE is defined as 
\begin{equation}\label{eq:LE}
    \nu(t) = \bra{\Psi_0} e^{i\hat{H}_f t} e^{-i\hat{H}_i t} \ket{\Psi_0},
\end{equation}
where $\ket{\Psi_0}$ is the initial state~\footnote{It is straightforward to see that it is obtained from the work characteristic function (\ref{eq:CharacteristicFunctionCorrelationFunction}), when the initial state is pure, i.e., $\hat{\rho}=|\Psi^{(i)}_0\rangle\!\langle\Psi^{(i)}_0|$.}.

In the first part of the experimental protocol, the system undergoes forward time evolution under the initial Hamiltonian $\hat{H}_i$ for a duration $t$, generating the state $e^{-i\hat{H}_i t}\ket{\Psi_0}$ (Fig.~\ref{fig:ExpProtocol_Full}, top panel). In phase space, this evolution entails the stretching of the Husimi distribution for the system dynamics, shown in the second snapshot of the middle panel.

The second part of the experimental protocol would need to implement the backward evolution $e^{i\hat{H}_f t}$. However, instead of implementing an actual time reversal that is experimentally challenging, we exploit the fact that the sign of the interaction parameter $U$ can be inverted through a Feshbach resonance, $U_f \rightarrow -U_f$~\cite{chin_feshbach_2010,Inouye1998}. Moreover, the tunnelling term $- K \hat{J}_z$ in the BJJ Hamiltonian is effectively reversed by applying two $\pi$-pulses around the $x$-axis:
\begin{equation}
    e^{i\pi \hat{J}_x} \hat{J}_z \, e^{-i\pi \hat{J}_x} = - \hat{J}_z \,.
\end{equation}
As illustrated in Fig.~\ref{fig:ExpProtocol_Full}, the evolution under $-U_f$ ($e^{i\hat{H}(-U_f) t}$ with $\hat{H}(-U_f) \equiv -U_{f} \hat{J}_x^2 - K \hat{J}_z$) is in-between the application of the two $\pi$-pulses, thus realizing an effective evolution generated by $-\hat{H}_f$ (backward evolution under $\hat{H}_f$). The corresponding evolution of the Husimi distribution is shown in the middle panel. While the experimental protocol partially recover the initial condition by refocusing the phase-space structure generated during the forward evolution, the final Husimi distribution does not coincide with the initial one because the backward dynamics is governed by $\hat{H}_f \neq \hat{H}_i$. 
This irreversibility directly reflects a finite value, different from unity, of the LE and therefore non-trivial work statistics induced by the interaction quench.

Therefore, the full protocol realizes the state 
\begin{equation}
    \ket{\Psi(t)} = e^{i\hat{H}_f t} e^{-i\hat{H}_i t} \ket{\Psi_0},
\end{equation}
whose overlap with the initial state yields the LE. The complex quantity $\nu(t)$ can be accessed experimentally using Ramsey interferometry schemes, where the real and imaginary parts of $\nu(t)$ are mapped onto measurable spin expectation values of an auxiliary qubit~\cite{SchemaInterferometrico}. 
The feasibility of the experimental protocol is demonstrated in the bottom panel of Fig.~\ref{fig:ExpProtocol_Full}. The LE obtained from the proposed experimental sequence (red crosses) perfectly coincides with the numerical value obtained from the ideal forward-backward evolution of \eqref{eq:LE} (solid blue line), thus validating the proposed implementation.  

\section{Conclusions}

In this paper, we have studied the quantum work distribution of a BJJ subjected to a sudden quench transformation, abruptly changing the self-interaction parameter. In our analysis, we have investigated different initial configurations of the junction, focusing on the conditions that lead to non-classical features of the work statistics.

As a first novel contribution, we have provided the analytical expression of the work probability distribution when the BJJ is prepared in the ground state before the application of the quench protocol. We have determined that the work distribution is a negative binomial with fractional index, and the corresponding work statistics depend only on the initial and final self-interaction parameters.

Then, we have expanded our analysis to include the possibility that the BJJ is prepared in the superposition state $|\Psi_0\rangle$ of the ground and second excited states of the BJJ undriven Hamiltonian $\hat{H}_{i}$. In order to properly take into account the presence of quantum coherence in $\hat{\rho} = |\Psi_0\rangle\!\langle\Psi_0|$, we have used the KDQs formalism that allowed us to describe the work distribution originated by the quench protocol of the self-interaction parameter. Interestingly, in such a case, we determined that the analytical expression of the work distribution is a weighted combination of negative binomial distributions. The corresponding work statistics depend again on the pre(post)-quench values of $U$, and exhibits non-positivity that varies depending on the specific initial superposition state. With the goal of exploiting such non-classical features, we have performed a numerical optimization over the complex coefficients of the superposition state (\ref{eq:SuperposTsate02}), finding the optimal values that maximizes the extractable work. Substituting these optimal values of the coefficients, the associated value of KDQs turned out to be real, and negative for some stochastic work $w_{j,k}$ values.
Comparing the KDQ work statistics with the statistics obtained by initializing the BJJ in the initial diagonal mixed state $\hat{\rho}_D$ with same populations of $\hat{\rho} = |\Psi_0\rangle\!\langle\Psi_0|$, we have shown how interference effects due to initial quantum coherences may lead to an increase of the extractable work and a decrease of the work variance.

We have also investigated the violation of a classical bound on the extractable work, which can be used as a witness of non-classical work statistics. We found a violation of such a bound for quenched BJJs initialized in the superposition state (\ref{eq:SuperposTsate02}), thus signalling the contextuality of the considered work protocols. 

Finally, we have proposed an experimental protocol to directly measure the characteristic function of the work distribution, in terms of a LE. Here, our main contribution is the implementation of the backward dynamics under $\hat{H}_{f}$ through Feshbach resonances and $\pi$-pulses.

In our analysis, all the analytical results and the experimental protocol are benchmarked through exact diagonalization (see App.~\ref{App:Exact diagonalization}). 

\subsection{Outlooks}

Among the various perspectives of this work, it is worth studying the application of a quench protocol to a BJJ with attractive interaction between the atoms, thus leading to a negative $U$. In such a case the system exhibits a quantum phase transition at $U=-\frac{K}{N}$, and it would be interesting to study the non-equilibrium thermodynamics of the BJJ at the transition. 

Another interesting direction is to investigate the role of initial coherence in finite-time work protocols. This would be the first step in designing quantum heat engines that use the BJJ as a work substance operating in finite time.

\section*{Acknowledgments}

The authors would like to thank Gabriele De Chiara for helpful discussions. This work has been financial supported by the PNRR MUR project PE0000023 NQSTI funded by the European Union---Next Generation EU, and by the European project ``High Performance Computer and Quantum Simulator hybrid (HPCQS)'' with grant agreement No.~101018180 within the European Union's Horizon 2020 research and innovation programme.

\appendix

\section{Numerical methods and validity of the HP approximation}\label{App:Exact diagonalization}

We here provide details of the exact diagonalization procedure, used in the numerical simulations, and we present a quantitative analysis of the validity of the HP approximation through comparison with exact many-body results. 

Throughout the paper, the total number of atoms in the condensate is assumed to be fixed and equal to $N$. Under this constraint, the Hilbert space associated to the two-mode Bose-Hubbard Hamiltonian (\ref{eq:BoseHubbard_Hamiltonian}) has dimension $(N+1)$. Consequently, once a basis is chosen, the state of the system can be described by a $(N+1)-$dimensional vector. For a fixed number of particle, a natural basis is given by the common eigenstates of the number operators $\hat{n}_s = \hat{c}^\dagger_s\hat{c}_s$ with $s=l,r$. Since $N$ is fixed, these states can be uniquely labelled with a single quantum number $n_l$ and denoted by $\ket{n_l}$. The physical observables are thus represented by square Hermitian matrices with finite dimension. For example, the Hamiltonian defined in \eqref{eq:BoseHubbard_Hamiltonian} assumes the tridiagonal form: 
\begin{equation}\label{eq:H_matrix}
\renewcommand{\arraystretch}{2.2} 
H = \begin{pmatrix}
    \dfrac{U}{4} N^2 & -\dfrac{K}{2} \sqrt{N} & 0 & \dots \\
    -\dfrac{K}{2} \sqrt{N} & \dfrac{U}{4} (N-2)^2 & -\dfrac{K}{2} \sqrt{2(N-1)} & \dots \\
    0 & -\dfrac{K}{2} \sqrt{2(N-1)} & \dfrac{U}{4} (N-4)^2 & \dots \\
    \vdots & \vdots & \vdots & \ddots
    \end{pmatrix}.
\end{equation}
The static and dynamic properties of the BJJ can be explored by means of exact diagonalization methods, for particle numbers relevant to typical experiments ($N<10^4$)~\cite{Gati2007}. In all numerical results presented in this work, the Hamiltonian (\ref{eq:H_matrix}) was diagonalized using the routine \texttt{eigh} from the \texttt{scipy.linalg} Python library. 

\begin{figure}[t]
    \includegraphics[width=0.9\linewidth]{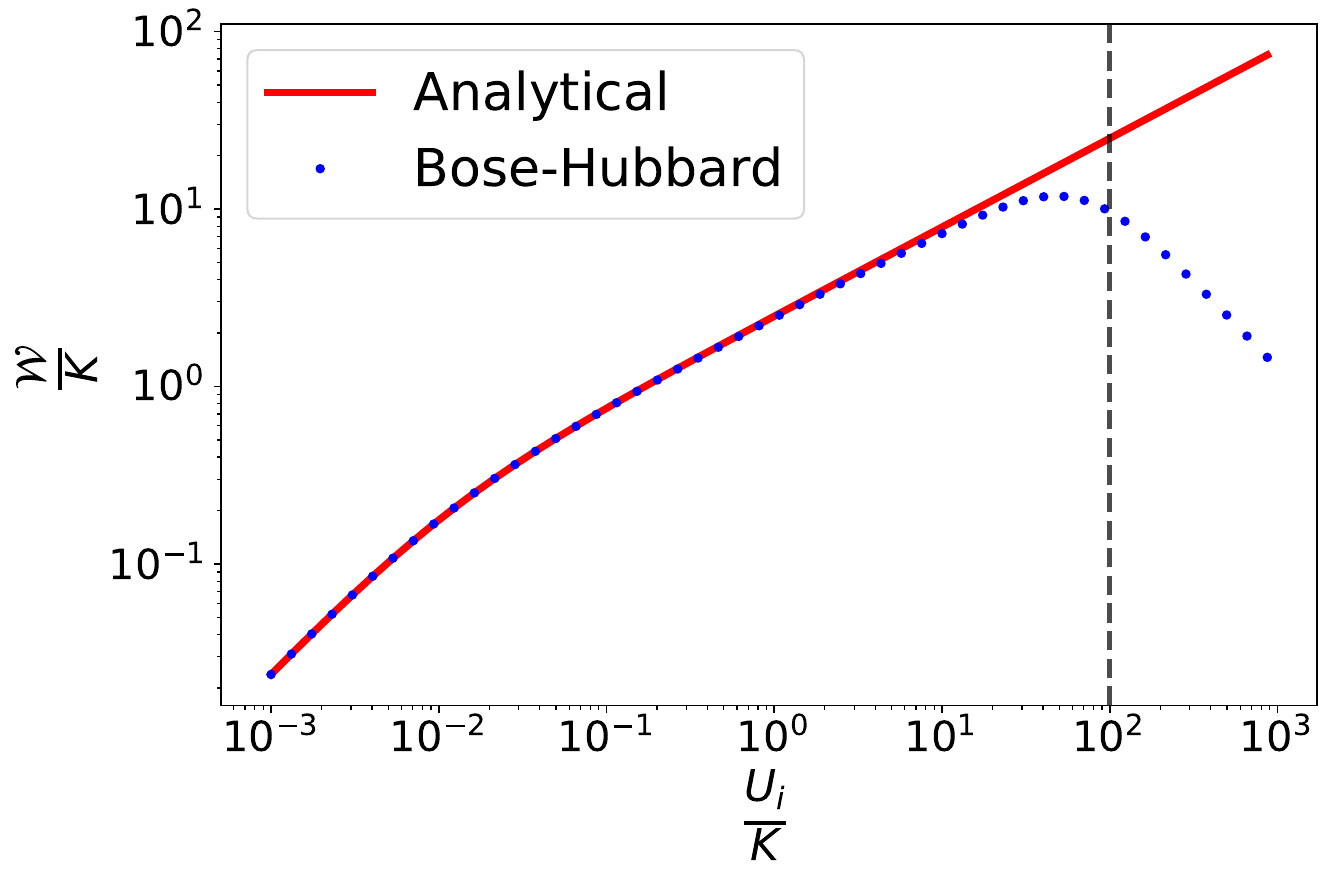}
    \includegraphics[width=0.9\linewidth]{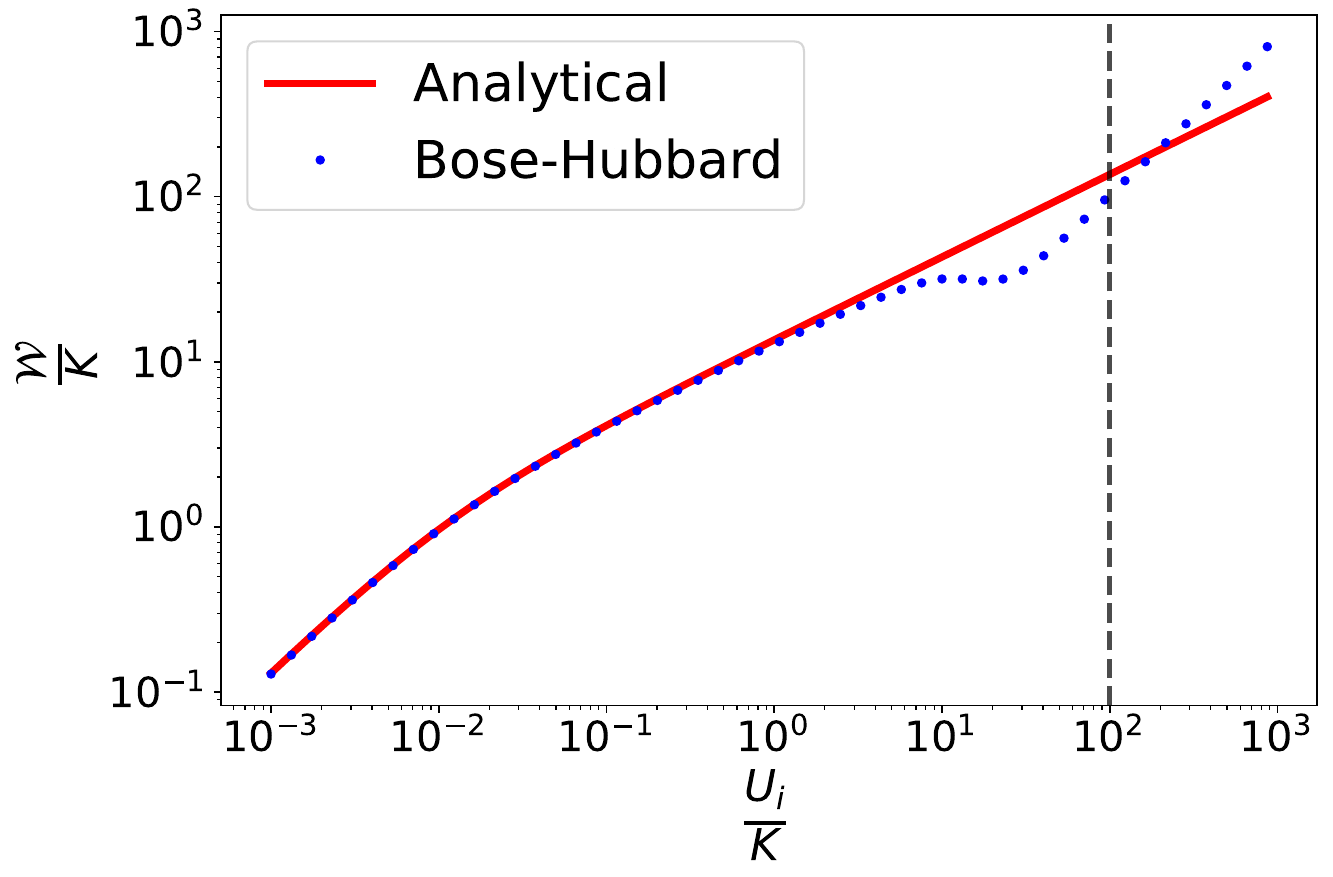}    
    \caption{Analytical (red solid line) and numerical (blue points) log-log plots of the average work $\mathcal{W}/K$ as a function of $U_i$, considering as initial state the ground state of the undriven BJJ Hamiltonian (top panel) and the superposition state (\ref{eq:SuperposTsate02}) (bottom panel), respectively. The plots are obtained by fixing $U_f=0$ and varying $U_i$ across all regimes of the BJJ, with $N=100$ atoms. In both panels, the black vertical dashed line corresponds to the theoretical upper boundary of the Josephson regime, beyond which the HP approximation breaks down.}
    \label{fig:HP_validity}
\end{figure}

As an illustration of the numerical technique described above, we now examine the range of validity of the HP approximation for the study of the work statistics of the BJJ. We consider a family of sudden quench protocols where the final self-interaction is fixed at $U_f=0$, while $U_i$ is varied across the Rabi, Josephson and Fock regimes. For each value of $U_i$, we consider two initial states of the BJJ: (i) the ground state and (ii) a superposition of the ground and second excited states, \eqref{eq:SuperposTsate02}, with coefficients $\alpha=0.303$ and $\beta=0.953$. A justification of the choice of this superposition state is given in App.~\ref{App: Optimization}. The numerical results shown in Fig.~\ref{fig:HP_validity} are obtained by first finding eigenvalues and eigenvectors of $\hat{H}_i$ and $\hat{H}_f$ and then evaluating the average work $\mathcal{W}=-\langle w\rangle$ by means of the definition in \eqref{eq:AverageWork_KD}.

In the case (i), the analytical predictions obtained using the HP approximation remain in good agreement with the exact numerical results until the ratio $\frac{U_i}{K}$ is of the order of $N$ (Fig.~\ref{fig:HP_validity}, top panel), which corresponds to enter the Fock regime. A different behaviour is observed for the superposition state [case (ii)]. In such a case, the agreement between the analytical and numerical extractable work is lost already when $\frac{U_i}{K}$ is of the order of unity (Fig.~\ref{fig:HP_validity}, bottom panel). Hence, the approximation is no longer valid before the system enters the Fock regime. The reason for this behaviour stems from the incapability of the HP approximation to describe the nonlinearity of the many-body energy spectrum. 

The fact that, for an initial superposition of Fock states, the disagreement emerges at lower values of $U_i$ compared to the case of an initial ground state, can be understood in the following way. The ground state of the many-body Hamiltonian is well-described by an effective QHO within the Rabi regime and for most of the Josephson regime. On the other side, the excited states can be well approximated by discretized QHO eigenfunctions only in the Rabi regime and in a smaller part of the Josephson regime. As the interaction is increased, thus approaching the Fock regime, this correspondence starts to fail. Consequently, when the initial state contains contributions from excited levels, the HP approximation deviates from the exact result at lower values of $U_i$ with respect to the ground-state case. This fact also better explains our choice to use the superposition of the two lowest energy levels of the undriven Hamiltonian as the initial state; see \eqref{eq:SuperposTsate02}. 

\section{Optimization of the initial state}
\label{App: Optimization} 

\begin{figure}[t]
    \includegraphics[width=0.9\linewidth]{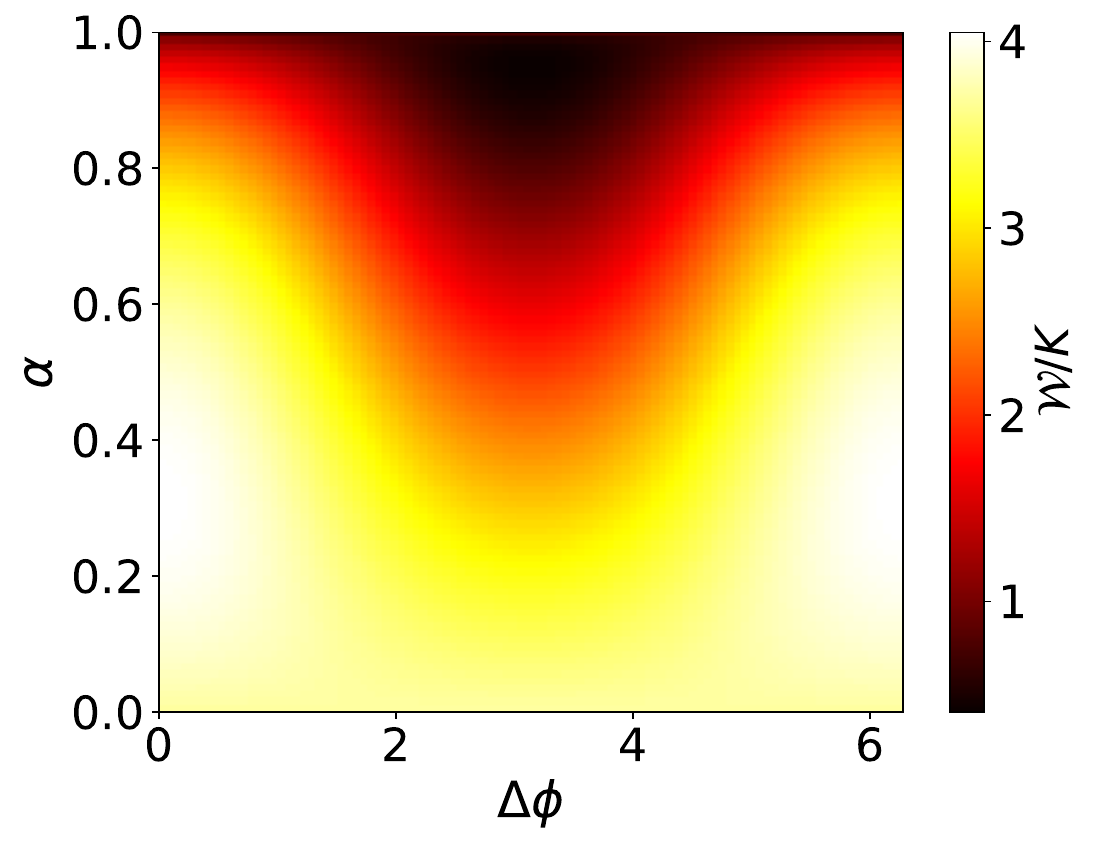}
    \includegraphics[width=0.9\linewidth]{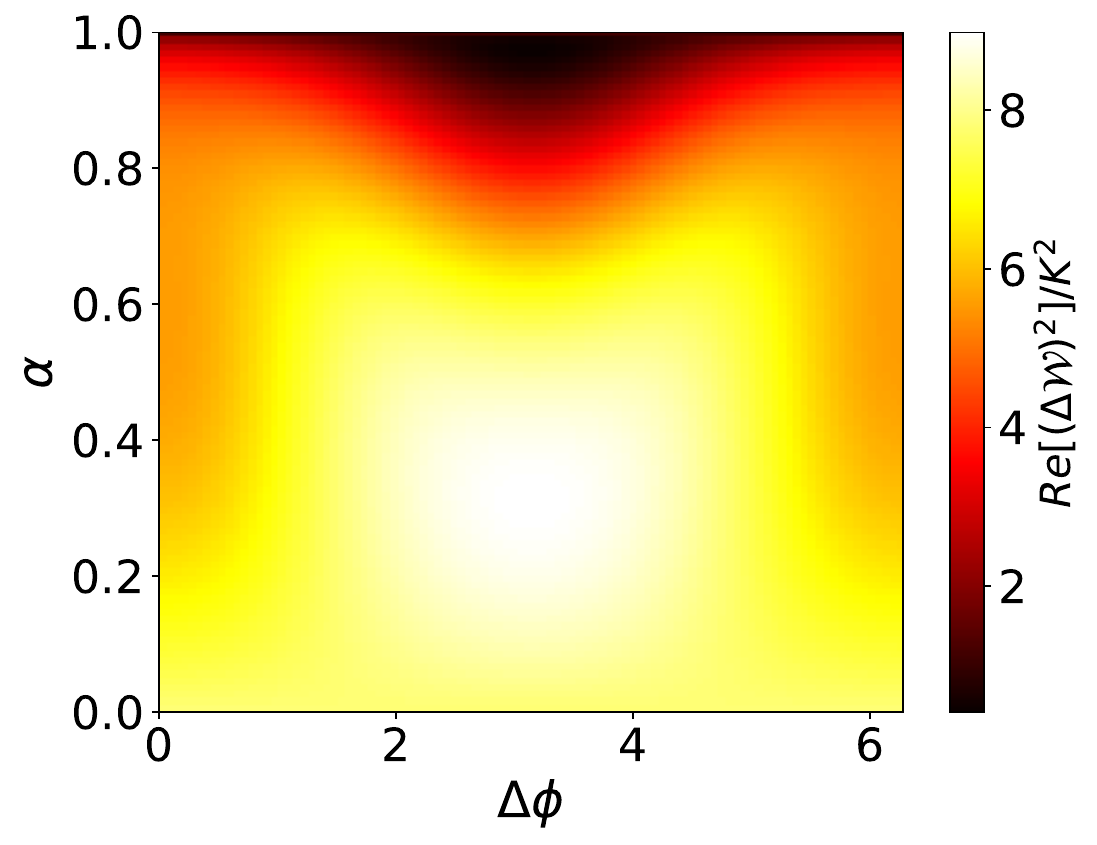}    
    \caption{Surface plot of the extractable work (top panel) and the real part of the work variance (bottom panel), as a function of $\alpha$ and of the relative phase $\Delta\phi$ between the coefficient of the superposition state (\ref{eq:SuperposTsate02}). The parameters of the quench protocol are $U_i=0.1K$ and $U_f=0K$.}
    \label{fig:ExtractableWorkSurfacePlot}
\end{figure}

In this Appendix we study the behaviour of the extractable work $\mathcal{W}$ as a function of the coefficients entering the superposition (\ref{eq:SuperposTsate02}) of the ground and second excited eigenstates of $\hat{H}_i$. The state (\ref{eq:SuperposTsate02}) can be parametrized by taking $\alpha \in \mathbb{R}$ and $\beta=\sqrt{1-\alpha^2}\,e^{i \Delta \phi}$, where $\Delta\phi$ is the phase difference between the two components of the superposition. Then, we consider a quench protocol from $U_i=0.1K$ to $U_f=0$. Using exact diagonalization (see App.~\ref{App:Exact diagonalization}), we evaluate $\mathcal{W}$ varying $\alpha \in [0,1]$ and $\Delta \phi \in [0,2\pi]$. The extractable work for the considered protocol is always positive (Fig.~\ref{fig:ExtractableWorkSurfacePlot}, top panel). The superposition state corresponding to the maximum extractable work is characterized by $\alpha_{\rm opt}=0.316$ and $\Delta\phi=0$, for which $\mathcal{W}^{\rm max}=4.05K$. 

Another relevant figure of merit in our analysis is the work variance. The superposition state that minimizes the real part of the work variance (Fig.~\ref{fig:ExtractableWorkSurfacePlot}, bottom panel) is characterized by $\alpha=0.98$ and $\Delta \phi=3.11$, for which $\mathcal{W}=0.414K$ and $\Re[(\Delta \mathcal{W})^2]=0.414K^2$. 
As expected, the state that minimizes the variance does not coincides with the one that maximizes the extractable work. For the latter state, one finds $\Re[(\Delta \mathcal{W})^2]=6.06K^2$, which is not the maximum value of the work variance. 

Finally, we remark that, since for the values of $U_i$ and $U_f$ considered here the HP approximation is valid, the same optimization procedure could have been done using the analytical expressions for the average work and the real part of the work variance reported in the main text. 

\section{Overlaps ratio}
\label{APP:TA_ratio}

Starting from the general expression for the overlaps of \eqref{eq:SqueezingFockLegendre}, we consider the ratio
\begin{equation}\label{eq:AppTA1}
    \frac{\Lambda_{2m,2}}{\Lambda_{2m,0}}=\sqrt{2}\,\text{sgn}\left(\omega_i-\omega_f\right) \frac{\mathcal{P}^{m-1}_{m+1}(x)}{\mathcal{P}^m_m(x)},
\end{equation}
where $\mathcal{P}^m_l(x)$ denotes the associated Legendre function~\cite{abramowitz1972handbook}. The ratio of Legendre functions can be simplified by using standard recurrence relations; in particular, 
\begin{subequations}
    \begin{align}
        &\sqrt{1-x^2}\,\mathcal{P}^{m}_{m+1}(x)=2x\mathcal{P}^{m-1}_{m+1}(x)-2m \mathcal{P}^{m-1}_m(x)\\
        &\mathcal{P}^m_{m+1}(x)=x\left(2m+1\right)\mathcal{P}^m_m(x)\\
        &\mathcal{P}^m_m(x)=-(2m-1)\sqrt{1-x^2}\,\mathcal{P}^{m-1}_{m-1}(x).
    \end{align}
\end{subequations}
Combining recursively these identities with the aim to eliminate the intermediate associated Legendre functions, one obtains:
\begin{equation}\label{eq:app_recursive}
    \frac{\mathcal{P}^{m-1}_{m+1}(x)}{\mathcal{P}^m_m(x)}=\frac{1-x^2(1+2m)}{2\sqrt{1-x^2}}.
\end{equation}
Substituting \eqref{eq:app_recursive} into \eqref{eq:AppTA1}, we determine \eqref{eq:AmplitudesRatio} in the main text.

\section*{Data Availability Statement}

All the codes used to produce Figs.~\ref{fig:NB_quench}-\ref{fig:ExtractableWorkSurfacePlot} are available at \cite{Orlandinigithub}.

\bibliography{bibliography}

\end{document}